\documentclass[aps,prl,twocolumn,showpacs,superscriptaddress]{revtex4-1}

\usepackage{longtable}
\usepackage{graphicx}
\usepackage{epsfig}
\usepackage{times}
\usepackage{amssymb}
\usepackage{color} 
\usepackage{times}
\usepackage{units}
\usepackage{textcomp}

\begin{document}

\title{Detection of 15\,dB Squeezed States of Light and their Application for the Absolute\\ Calibration of Photoelectric Quantum Efficiency}

\author{Henning Vahlbruch$^\ast$}
\affiliation{Institut f\"ur Gravitationsphysik, Leibniz Universit\"at Hannover and Max-Planck-Institut f\"ur Gravitationsphysik (Albert-Einstein-Institut), Callinstr. 38, 30167 Hannover, Germany}
\author{Moritz Mehmet}
\affiliation{Institut f\"ur Gravitationsphysik, Leibniz Universit\"at Hannover and Max-Planck-Institut f\"ur Gravitationsphysik (Albert-Einstein-Institut), Callinstr. 38, 30167 Hannover, Germany}
\author{Karsten Danzmann}
\affiliation{Institut f\"ur Gravitationsphysik, Leibniz Universit\"at Hannover and Max-Planck-Institut f\"ur Gravitationsphysik (Albert-Einstein-Institut), Callinstr. 38, 30167 Hannover, Germany}
\author{Roman Schnabel}
\affiliation{Institut f\"ur Gravitationsphysik, Leibniz Universit\"at Hannover and Max-Planck-Institut f\"ur Gravitationsphysik (Albert-Einstein-Institut), Callinstr. 38, 30167 Hannover, Germany}
\affiliation{Institut f\"ur Laserphysik and Zentrum f\"ur Optische Quantentechnologien, Universit\"at Hamburg, Luruper Chaussee 149, 22761 Hamburg, Germany}

\date{15 June 2016 / 29 March 2026}

\begin{abstract}
Squeezed states of light belong to the most prominent nonclassical resources. They have compelling applications in metrology, which has been demonstrated by their routine exploitation for improving the sensitivity of a gravitational-wave detector since 2010. Here, we report on the direct measurement of 15 dB squeezed vacuum states of light and their application to calibrate the quantum efficiency of photoelectric detection. The object of calibration is a customized InGaAs positive intrinsic negative ({\it p-i-n}) photodiode optimized for high external quantum efficiency. The calibration yields a value of 99.5\,\% with a 0.5\% ($k = 2$) uncertainty for a photon flux of the order $10^{17}\,$s$^{-1}$ at a wavelength of 1064nm. The calibration neither requires any standard nor knowledge of the incident light power and thus represents a valuable application of squeezed states of light in quantum metrology.
\end{abstract}

\maketitle

Squeezed light was first produced by four-wave mixing in sodium atoms in 1985 \cite{SHYMV85} and shortly after by degenerate parametric down-conversion in a nonlinear crystal placed inside an optical resonator, also called optical parametric amplification \cite{Wu1986}. Since then, the latter approach has been used in many squeezed-light experiments, as reported for instance in \cite{Schneider1998, Furusawa1998, Bowen2003tele, Lam1999, Aoki2006, Vahlbruch2008, Eberle2010, Mehmet2011, Stefszky12}. The down-converted photon pairs exiting the squeezing resonator show quantum correlations that lead to a squeezed photon-counting noise upon detection. The closer the squeezing resonator is operated to its oscillation threshold (thr) the stronger the quantum correlations are and thus the larger the generated squeezing is. In principle, a squeezed-light source can produce an infinite squeezing level; however, the measured squeezing level is usually limited by photon loss during squeezed-light generation, propagation, and detection. 
Also, phase noise \cite{Aoki2006}, excess noise \cite{Bowen2002} and detector dark noise \cite{Schneider1998} have been found to impair the observable squeezing strength. 
An important milestone regarding squeezing strength was achieved in 2008 when a 10 dB squeezed vacuum state of light was measured for the first time \cite{Vahlbruch2008}. In subsequent experiments the detected squeezing levels could be further increased, leading to values of up to 12.7\,dB \cite{Eberle2010, Mehmet2011}. 

Squeezed states of light allow for fundamental research on quantum physics, e.g.~on the famous Einstein-Podolsky-Rosen-paradox \cite{Einstein1935,Ou1992,Haendchen2012} and they have been proposed for compelling applications for quantum enhanced metrology \cite{Caves1981,Caves1982}. 
The advanced LIGO detectors, during the recent detection of gravitational waves from a binary black hole merger \cite{LSC2016, Abbott2016}, operated at a sensitivity mainly limited by quantum noise. To further increase the sensitivity a reduction of quantum noise is obligatory. 
Since 2010, squeezed states of light have been routinely used to increase the quantum noise limited sensitivity of the gravitational-wave (GW) detector GEO\,600 \cite{LSC2011,Grote2013}. 
This first true application and the demonstration of a squeezing enhanced LIGO detector \cite{Aasi2013} support the plans for future GW detectors, such as the Einstein Telescope \cite{Punturo2010} and upgrades to advanced LIGO, to include the squeezed light injection technique.

A 10\,dB quantum enhancement of future GW detectors seems to be within reach, but requires a total photon loss of less than 10\% for the squeezed field. This includes all losses that occur during the generation of the squeezed field, its injection into and propagation through the GW detector, and finally the photoelectric detection. In contrast to the loss of optical components the photoelectric quantum efficiency of a positive intrinsic negative ({\it p-i-n}) photodiode is more difficult to determine. To calibrate the quantum efficiency the photon flux of the light beam needs to be known or a calibrated detector is required for comparison. With the exploitation of radiation standards, measurements of absolute external quantum efficiencies with measurement uncertainties down to  to 0.5 $\times$ 10$^{-3}$ have been achieved at some near-infrared wavelengths for optical radiant powers of up to 400 $\mu$W \cite{Lopez2006,Eppeldauer2009}. 

For 1050 nm, measure- ment uncertainties between 0.1\% and 0.15\% were achieved in an international comparison \cite{Brown2010}, carried out under incoherent irradiation with optical radiant powers between 1 and 100\,$\mu$W. For laser power in the mW range, radiation standards provide a measurement uncertainty at 1064 nm of around 0.2\% \cite{BIPM}. For the regime of photon-number- resolving photodetectors quantum correlations of single- photon pairs generated via parametric down-conversion have been used as an absolute calibration technique, for example \cite{Migdall1999, Brida2006, Avella2011,Avella2016}, and accuracies at the level of parts in 10$^3$ have been achieved \cite{Polyakov2007}.

Here, we present the realization of a novel approach for the precise calibration of absolute external quantum efficiencies of {\it p-i-n} photodiodes based on a continuous-wave squeezed-light source. Our method does not require any calibrated standard for the incident light power, but only the measurements of squeezing and corresponding antisqueezing levels and a determination of optical loss of components. For this new application of quantum metrology we have realized a low-loss squeezed-light experiment, which allowed for the direct observation of up 15 dB squeezing. This is the strongest quantum noise reduction demonstrated to date and enabled the low error bar of our method.

We calibrated the quantum efficiency of an InGaAs {\it p-i-n} photo diode to 99.5\% with 0.5\% ($k=2$) uncertainty. The device under test was a custom-made photodiode manu- factured from the same wafer material as the photodiode used in GEO 600 since 2011 \cite{LSC2011}.

\begin{figure}[t!]
\centerline{
\includegraphics[width=87mm]{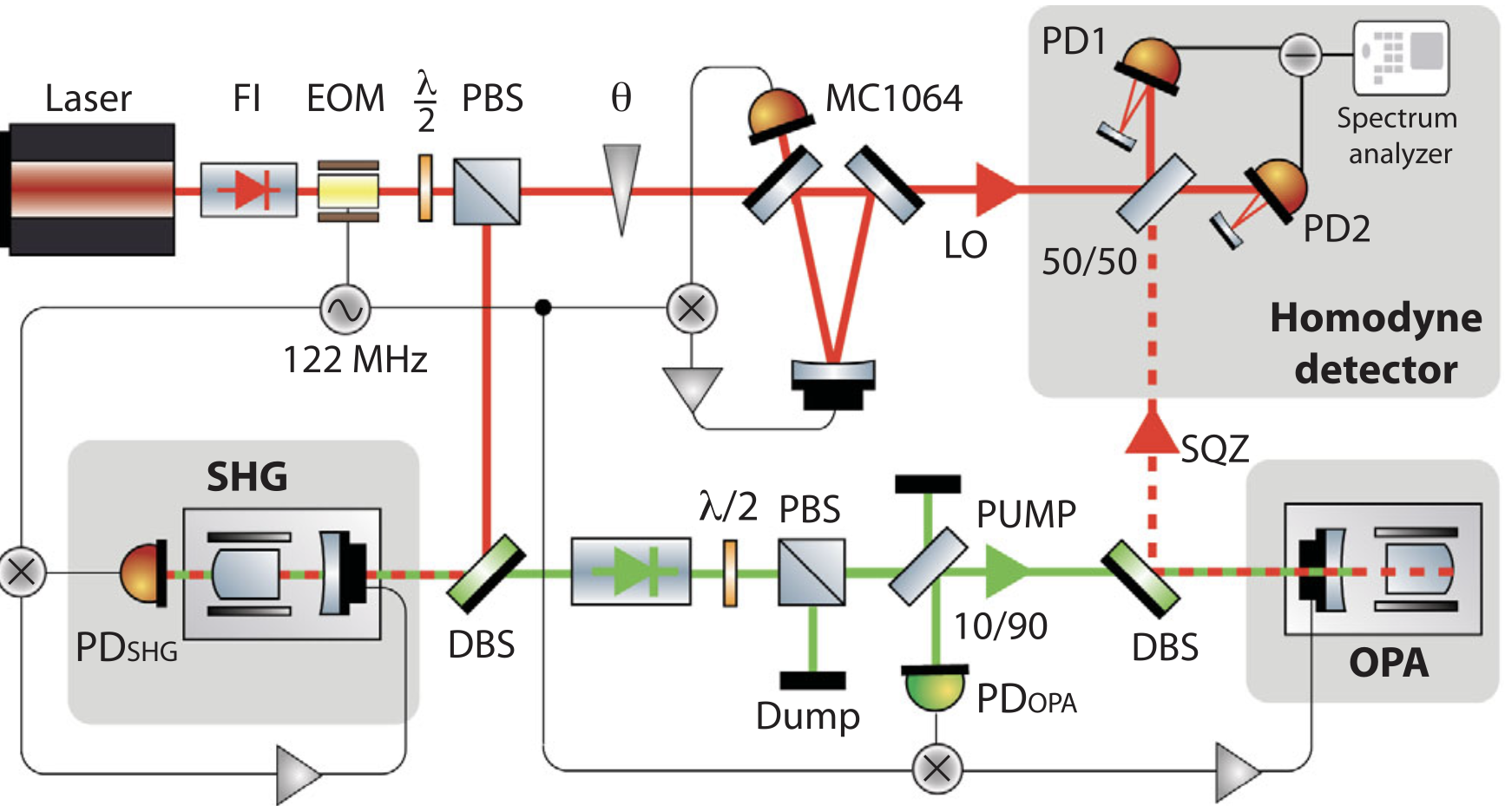}}
  \vspace{-2mm}
\caption{Schematic of the experimental setup. Squeezed vacuum states of light (SQZ) at a wavelength of 1064\,nm were generated in a doubly resonant, type\,I optical parametric amplifier (OPA) operated below threshold. 
SHG: second harmonic generation, PBS: polarizing beam splitter;  DBS: dichroic beam splitter; LO: local oscillator, PD: photodiode; MC1064: three-mirror ring cavity for spatiotemporal mode cleaning; EOM: electro-optical modulator; FI: Faraday Isolator. The phase shifter for the relative phase $\theta$ between SQZ and LO was a piezo actuated mirror. Retro reflecting mirrors were used to recycle the residual reflection from the photo diodes used for homodyne detection. An auxiliary beam (not shown) was used for the alignment of the homodyne contrast.}
  \label{fig:1}
\end{figure}

The schematic of our experimental setup is illustrated in Fig.\,\ref{fig:1}. The laser source was a monolithic non-planar Nd:YAG ring laser with 2\,W of continuous-wave single-mode output power at a wavelength of 1064\,nm.  About 350\,mW were converted to approximately 180\,mW at 532\,nm in a second harmonic generation unit. 
The second-harmonic light was used to pump a doubly resonant optical parametric amplifier (OPA) in which squeezed vacuum states were generated by 
type\,I degenerate parametric down-conversion (PDC). A standing-wave cavity was formed by an external coupling mirror and the back surface of a 9.3\,mm long periodically poled KTP (PPKTP) crystal. The crystal had a 10\,mm radius of curvature (roc) and was dielectrically coated for a high reflectivity (HR) for the fundamental (R\,=\,99.96($^{+0.01}_{-0.01}$)\,\% at 1064\,nm) and the pump field (R\,$>$\,99.9\% at 532\,nm). The plane crystal front face was anti-reflectively (AR) coated for both wavelength. The coupling mirror (roc\,=\,25\,mm) had a power reflectivity of R\,=\,97.5($^{+0.5}_{-0.5}$)\,\% at 532\,nm and R\,=\,87.5\,($^{+0.5}_{-0.5}$)\,\% at 1064\,nm. 
The cavity length was approximately 40\,mm, resulting in a free spectral range of 3.75\,GHz,  a finesse of 243 at 532\,nm (linewidth\,= \,15.4\,MHz) and a finesse of 54 at 1064\,nm (linewidth\,=\,69.4\,MHz). 
The cavity length was held on resonance for the fundamental and pump laser fields via a Pound-Drever-Hall locking scheme relying on the detection of the green pump field. To generate the control signal 10\,\% of the pump field was detected with a photo detector (PD$_{\mathrm{OPA}}$) via a beam splitter as shown in Fig.\,\ref{fig:1}.
The required phase modulation was generated with an electro-optical modulator (EOM) driven at 122\,MHz. The squeezed light at a wavelength of 1064\,nm was separated from the pump field at 532\,nm using a dichroic beamsplitter (DBS).\\

The measurements of the vacuum noise, and the squeezed and anti-squeezed quantum noise, were performed by means of balanced homodyne detection. As sketched in Fig.~\ref{fig:1}, for this detection scheme two input fields (LO and signal) need to interfere at the 50-50 beamsplitter. Each of the resulting two output fields was detected with one  {\it p-i-n} photo diode (custom made by Laser Components). The diodes had an active area of 500\,$\mu$m diameter and were AR-coated for a 20{\textdegree} angle of incidence. 
A signal proportional to the difference of their photo currents was generated, which was then fed into a spectrum analyzer. This yields the variance of a generic quadrature  $\hat X(\theta)= \hat X_{1} \cos \theta   + \hat X_{2} \sin \theta $, where $\theta$ is the phase angle between the local oscillator and the signal field. Consequently, the phase angle defines the quadrature under investigation. 
We used a piezo actuated mirror as the phase shifter to adjust $\theta$. The potential power fluctuations of the local oscillator field due to beam pointing with respect to the eigenmode of MC1064 were found to be negligible over the necessary actuation range. We achieved a fringe visibility in the TEM$_{00}$ mode of 99.6\,($^{+0.05}_{-0.05}$)\,\% between the signal input and the local oscillator beam. To recycle the residual reflection from each photo diode surface (which we determined to be 0.3\,\%) two curved HR mirror were used as a retro reflectors. For all measurements presented here a local oscillator intensity of 26.5\,mW  corresponding to a photon flux of $1.4\cdot10^{17}\,$s$^{-1}$ was used. 
This resulted in a clearance between the electronic detection dark noise and the vacuum noise reference level of up to 28\,dB at Fourier-frequencies between 3 and 10\,MHz.\\   
%
%
%
%

\begin{figure}[t!]
  \centerline{\includegraphics[width=87mm]{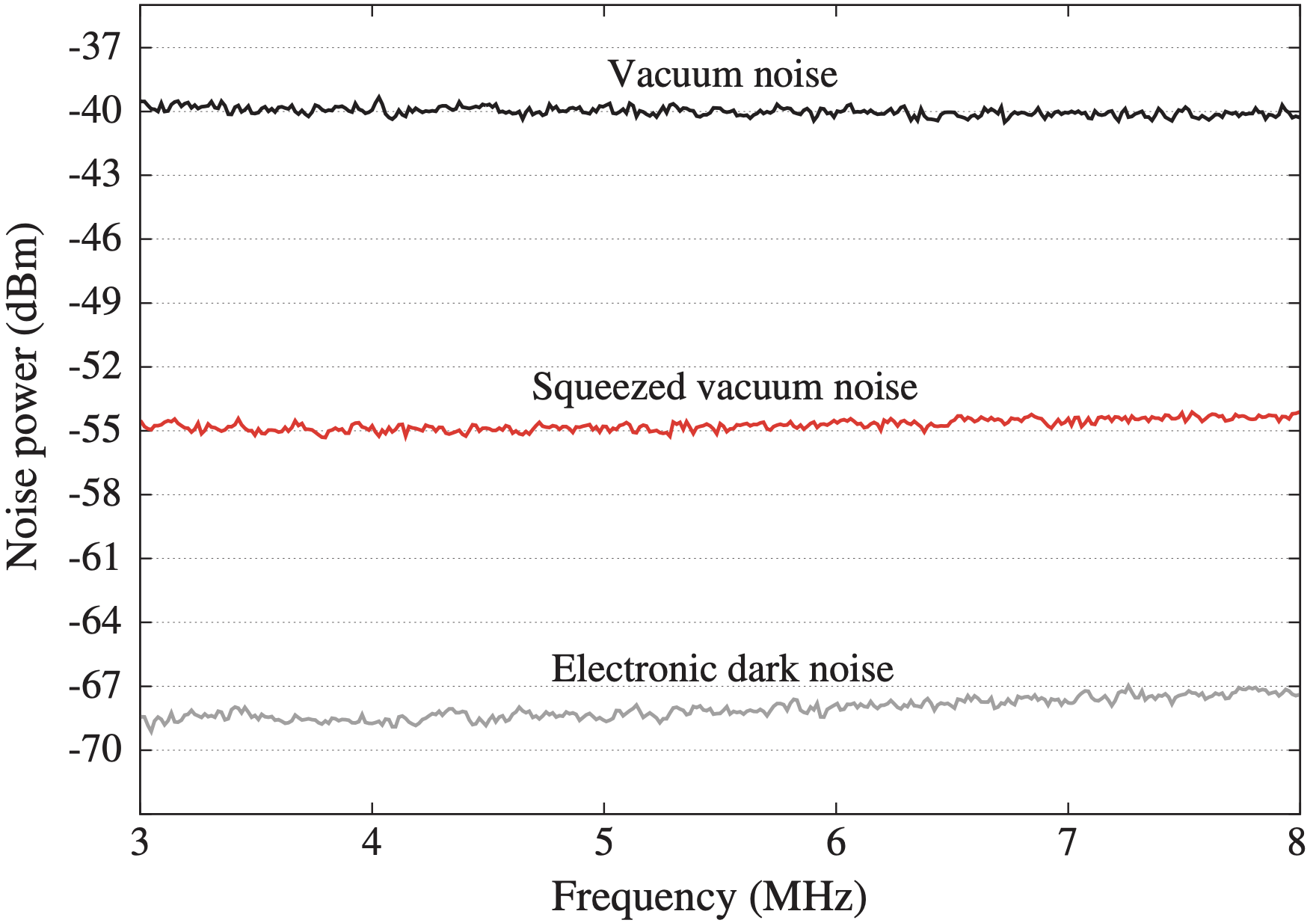}}
  \vspace{-2mm}
\caption{Quantum noise and electronic dark noise at Fourier frequencies from 3 to 8\,MHz. The vacuum noise reference level corresponds to local oscillator power of 26.5\,mW. The measurement time for each individual trace was 211\,ms. A non-classical noise reduction of up to 15\,dB below vacuum noise was directly observed. The degradation of the squeezing factor towards higher frequencies is a consequence of the OPA line-width. The electronic detector dark noise was not subtracted from the data.}
  \label{fig:2}
\end{figure}

The measurements in the  frequency range from 3\,MHz to 8\,MHz, as shown in Fig.~\ref{fig:2}, were recorded with a Rohde\&Schwarz FSU Spectrum Analyzer with a resolution bandwidth (RBW) of 300\,kHz and a video bandwidth (VBW) of 200\,Hz. The electronic dark noise of the detection system was recorded with the signal and local oscillator input blocked. The vacuum noise was recorded with only the local oscillator port of the balanced beam splitter open. In this configurations no photons entered through the signal port such that the measured noise directly represents the vacuum noise reference level. The squeezed noise was recorded by additionally opening the signal input port and adjusting the phase between local oscillator and signal field such that the measured noise variance was minimized. Up to 15\,dB squeezing was measured with merely 16\,mW of second harmonic pump power. In a separate measurement we confirmed the linearity of the detection system, including the spectrum analyzer, by measuring shot-noise levels versus local oscillator powers.

For a detailed analysis we compared the performance of our apparatus with a theoretical model. 
The output spectrum of the squeezed ($-$) and anti-squeezed ($+$) quadrature variances for an OPA below threshold can be computed as \cite{Polzik1992} 
\begin{equation}\label{sqzspec}
 	\Delta^2 \hat X_{+,-} =  1\pm \eta \frac{4 \sqrt{P/P_{\mathrm{thr}}}} {\left(1\mp \sqrt{P/P_{\mathrm{thr}}}\right)^2+4 \left(  \frac{2 \pi f}{\gamma} \right  )^2}\ ,
\end{equation}
where $\eta_{\rm tot}$ is the total detection efficiency ($\eta_{\rm tot} = 1 -$ total optical loss), $P$ is the second-harmonic pump power, $P_{\mathrm{thr}}$ is the amount of pump power required to reach the threshold of OPA, and $f$ is the sideband frequency of the measurement. The cavity decay rate $\gamma=c(T+L)/l$, with the speed of light $c$, the cavity round-trip length $l$, the coupling mirror's power transmissivity $T$, and the round trip loss $L$. In addition to optical loss, one needs to include the impact of phase fluctuations between the signal and LO field.
While the effect of optical loss is the addition of contributions from the unsqueezed vacuum field to the squeezed quadrature, the effect of phase noise is to add extra noise through components proportional to the antisqueezed quadrature. Assuming that potential phase fluctuations follow a Gaussian distribution and that the standard deviation is small, an rms phase jitter $\theta_{\mathrm{pn}}$ corresponds to the homodyne detector measuring at an offset phase angle $\theta_{\mathrm{pn}}$ relative to the ideal quadrature which yields variances of the form \cite{Aoki2006}  
\begin{equation}\label{sqzspecPN}
V_{+,-} =  \Delta^2 \hat X_{+,-} \cos^2({\theta_{\mathrm{pn}}}) +  \Delta^2 \hat X_{-,+} \sin^2({\theta_{\mathrm{pn}}})\ .
\end{equation}
\begin{figure}[t!]
  \centerline{\includegraphics[width=87mm]{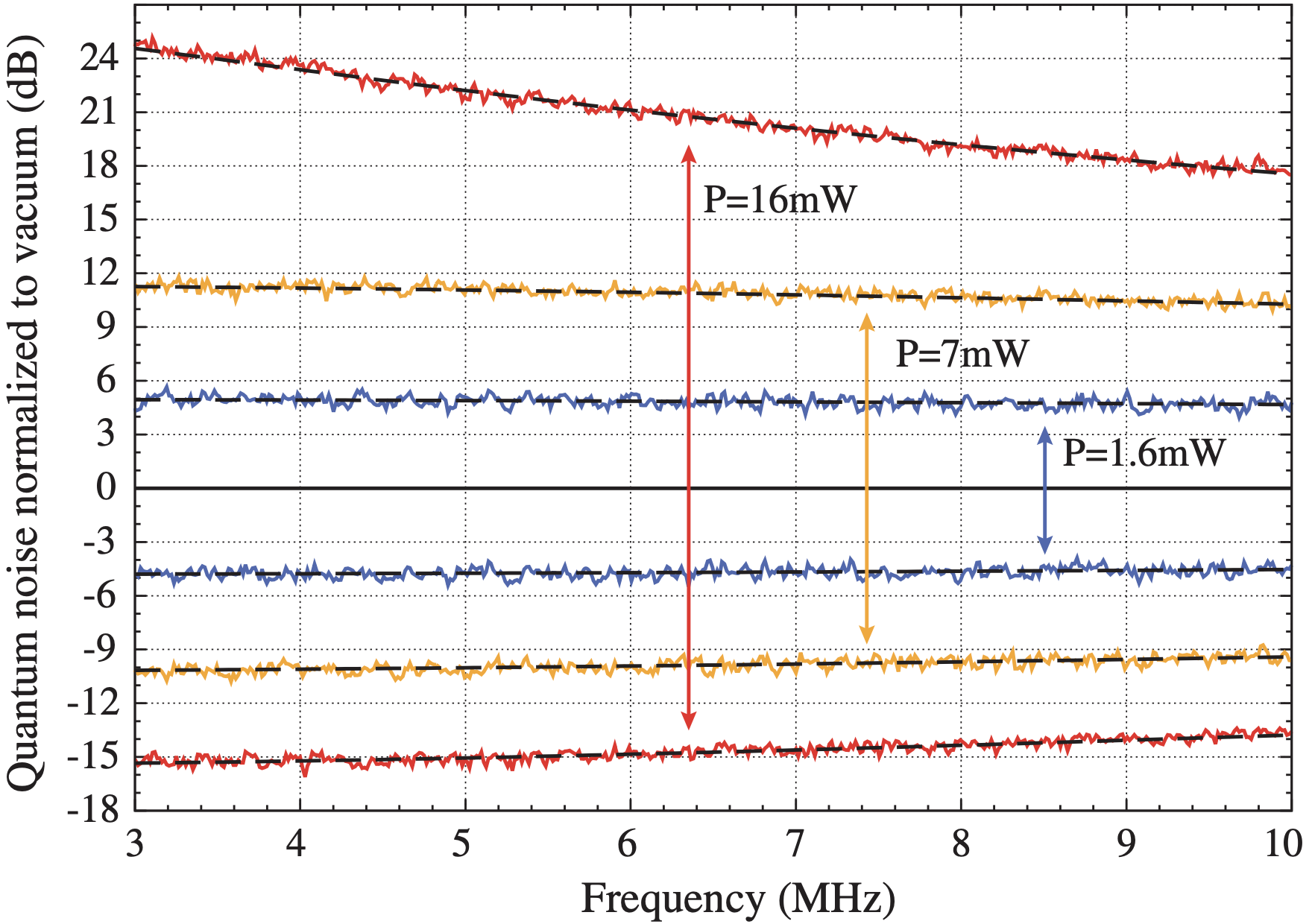}}
  \vspace{-2mm}
\caption{Pump power dependence of the squeezing and anti-squeezing spectra, experiment and theory. The theoretical curves (dashed lines) were modeled with $\eta=0.975$ and $\theta_{\textrm{pn}}=1.7$\,mrad for pump powers corresponding to $P/P_{\mathrm{thr}}$ of 8\%, 33.9\%, and 83.5\% and a full-width-half-maximum linewidth $\gamma / 2\pi =$ 84\,MHz. The electronic dark noise was subtracted from the data, which were subsequently normalized to the vacuum noise level. The measurement time for each individual trace was 295\,ms. Already with a pump power of 1.6\,mW, a nonclassical noise reduction of up to 5 dB was obtained. With 7\,mW our setup produced up to 10\,dB squeezing with only 11\,dB antisqueezing. The model is in good agreement with the 0.3\,dB increase to 15.3\,dB of maximum squeezing due to subtraction of the dark noise. This effect is negligible for all other traces in this representation.}
\label{fig:3}
\end{figure}
This model was used to fit the spectral distributions shown in Fig.~\ref{fig:3}.
These measurements are squeezing and anti-squeezing spectra obtained with second harmonic pump powers of 1.6\,mW, 7\,mW, and 16\,mW, respectively.  
Here, the electronic dark noise was subtracted from the measured data (anti-squeezing, vacuum, and squeezing) and subsequently all traces were renormalized to the (new) vacuum noise level. 
From the squeezing level of 15.3\,dB shown here, we deduce that the directly measured squeezing as shown in Fig.~\ref{fig:2} was degraded by 0.3\,dB due to contributions by electronic darknoise. The dashed lines in Fig.~\ref{fig:3} correspond to the above model with ratios $\nicefrac{P}{P_{\mathrm{thr}}}$ of 8\%, 33.9\%, and 83.5\% with a total detection efficiency $\eta=0.975$ and a phase noise $\theta_{\mathrm{pn}}=1.7$\,mrad.

The detected squeezing level is not limited by phase noise but by 2.5\,($^{+0.1}_{-0.1}$)\,\% optical loss, of which we attribute 0.8\,($^{+0.1}_{-0.1}$)\,\% to the homodyne contrast (99.6\,($^{+0.05}_{-0.05}$)\,\%) and 0.2\,($^{+0.01}_{-0.01}$)\,\% to the transmission loss through lenses. The latter we measured in a separate experiment.   
We determined the OPA escape efficiency to $\eta_{\mathrm{esc}}=\nicefrac{T}{(T+L)}=99.05(^{+0.4}_{-0.45})$\,\%. For this we take into account the loss of the PPKTP crystal comprising the residual transmission through the HR-coated backside 400$(^{+100}_{-100}$)\,ppm, the (negligible) absorption within the crystal (12\,ppm/cm, \cite{MeierPhD, Bogan2015}, the residual reflection of the AR-coated frontside (two times 400$(^{+200}_{-200}$)\,ppm, and the coupling mirror transmissivity $T=0.125(^{+0.005}_{-0.005})$. This analysis accounts for $2.0(^{+0.5}_{-0.6})$\,\% of the 2.5\,($^{+0.1}_{-0.1}$)\,\% overall optical loss in our squeezing path. The remaining contribution to the loss budget is the quantum efficiency of the photo diodes, which we therefore deduce to 99.5$(^{ {+0.5} }_{-0.5})$\,\%. 
%
%
%

In conclusion, we have realized a low-loss and low phase noise squeezed-light experiment with a doubly resonant, nonmonolithic OPA cavity. From a comparison to a theoretical model we determined the total optical loss to be 2.5\,($^{+0.1}_{-0.1}$)\,\% and the upper bound for phase noise to be 1.7\,mrad.
With only 7\,mW pump power 10\,dB squeezing with 11\,dB anti-squeezing were measured which are the purest strongly squeezed states observed to date. 
The generation of the high squeezing factor (10\,dB), together with a low anti-squeezing factor (11\,dB), which is close to the lower bound set by Heisenberg's uncertainty relation, is of high relevance for the application in GW detectors. Low anti-squeezing is important to minimize back-action noise and to make back-action evading (quantum non-demolition) schemes, e.g. that one in \cite{Jaekel1990}, more efficient. By the implementation of a scheme for the coherent control of audioband squeezed vacuum states  \cite{Vahlbruch2006} our apparatus can be extended to be compatible with the application in gravitational-wave detectors. 
With an increased pump power up to 15\,dB squeezing was directly measurable for the first time.
These states were used for an absolute calibration of the photo- electric external quantum efficiency of a custom-made {\it p-i-n} InGaAs photodiode to 99.5\% with 0.5\% ($k = 2$) uncertainty at a wavelength of 1064\,nm. Our method does not rely on the measurement of absolute optical power, as realized for example in \cite{Lopez2006} by means of an electrically calibrated cryogenic radiometer (ECCR). 
Since no reference detector is required for comparison nor photon flux needed to be known, it is rather a direct measurement of the external quantum efficiency. The high precision documents the potential of this novel application of squeezed states in quantum metrology. Any additional reduction of uncertainties, especially of the OPA escape efficiency, will directly enhance the accuracy of this calibration method even further.\\[-3mm]

The work was supported by RS's core funding from Leibniz Universit\"at Hannover. 
HV acknowledges the Deutsche Forschungsgemeinschaft (DFG) in the context of project VA 1031/1-1 (Sept. 2015 -- Aug. 2018), which funded his position during the final phase of the project, 
when the data was prepared for publication and the manuscript was written.

\end{document}